\documentclass{article}

\usepackage{amssymb}
\usepackage{graphicx}
\usepackage{psfrag}
\usepackage{amsfonts}
\usepackage{latexsym}
\usepackage{amsmath}

\newtheorem{definition}{Definition}
\newtheorem{lemma}{Lemma}

\newtheorem{theorem}{Theorem}

\def\hal{\vrule height 5 pt width 0.05 in  depth 0.8 pt}
\def\ms{{\medskip }}

\newcommand{\fin}{\hfill \hal \ms}

\newcommand{\tr}{{\triangle}}
\newcommand{\dem}{\noindent {\bf Proof: }}

\title{Minimal Convex Decompositions}

\author{M. Lomel\'{\i}-Haro\\
Instituto de F\'{\i}sica-Universidad Aut\'onoma de San Luis Potos\'{\i}\\
lomeli@ifisica.uaslp.mx}

\begin{document}

\maketitle

\begin{abstract}
Let $P$ be a set of $n$ points on the plane in general position. We say that a set $\Gamma$ of convex polygons with vertices in $P$ is a convex decomposition of $P$ if: Union of all elements in $\Gamma$ is the convex hull of $P,$ every element in $\Gamma$ is empty, and for any two different elements of $\Gamma$ their interiors are disjoint. A minimal convex decomposition of $P$ is a convex decomposition $\Gamma'$ such that for  any two adjacent elements in $\Gamma'$ its union is a non convex polygon. It is known that $P$ always has a minimal convex decomposition with at most
$\frac{3n}{2}$ elements. Here we prove that $P$ always has a minimal convex decomposition with at most
$\frac{10n}{7}$ elements.
\end{abstract}

\section{Introduction}

Let $P_n$ denote a set of $n$ points on the plane in general position. We denote as $Conv(P_n)$ the convex
hull of $P_n$ and $c$ the number of its vertices, and given a polygon $\alpha$ we denote as $\alpha^o$ its interior. We say that a set
$\Gamma=\{\gamma_1,\gamma_2,...,\gamma_k\}$ of $k$ convex polygons with vertices in $P_n$ is a {\em convex decomposition} of $P_n$ if:

(C1) Every $\gamma_i \in \Gamma$ is empty, that is, $P_n \cap \gamma_i^o = \emptyset$ for $i=1,2,...,k.$

(C2) For every two different $\gamma_i,$ $\gamma_j \in \Gamma,$ $\gamma_i^o \cap \gamma_j^o = \emptyset.$

(C3) $\gamma_1 \cup \gamma_2 \cup ... \cup \gamma_k = Conv(P_n)$.

In \cite{openproblems} they conjectured that for every $P_n$ there is a convex decomposition with at most $n+1$ elements. This was disproved in \cite{aichholzer} giving an $n$--point set such that every convex decomposition has at least $n+2$ elements. Later in this direction, in \cite{garcia} 
they give a point set $P_n$ on which every convex decomposition has at least $\frac{11n}{10}$ elements.

We are interested in convex decompositions of $P_n$ with as few elements as possible. A {\em triangulation} of $P_n$ is a convex decomposition $T = \{t_1,t_2,...,t_k\}$ on which every $t_i$ is a triangle. In \cite{simflippingedges} they prove that any triangulation $T$ of $P_n,$ has a set $F$ of at least
$\frac{n}{6}$ edges that, by removing them we obtain $|F|$ convex quadrilaterals with disjoint interiors.  So $\Gamma = T \setminus F$ is a convex decomposition yielding the bound $|\Gamma| \leq \frac{11n}{6}-c-2.$ We have the following definition.

\begin{definition}
Let $\Gamma$ be a convex decomposition of $P_n.$ If the union of any two different elements in $\Gamma$  is a nonconvex polygon, then $\Gamma$ will be called {\em minimal convex decomposition.}
\end{definition}

In \cite{descconv} they show that any given set $P_n$ always has a minimal convex decomposition with at most $\frac{3n}{2}-c$ elements. Here we improve this bound giving a minimal convex decomposition of $P_n$ with at most $\frac{10n}{7}-c$ elements.

\section{Minimal Convex Decompositions}

Let  $p_1=(x_1,y_1)$ be the element in $P_n$ with the lowest $y$--coordinate.  If there are two points with same $y$--coordinate we take $p_1$ as the element with the smallest $x$--coordinate.

\begin{figure}[h]
\centering

\psfrag{pn}{$p_n$}
\psfrag{pn-1}{$p_{n-1}$}
\psfrag{p1}{$p_1$}
\psfrag{p2}{$p_2$}
\psfrag{p3}{$p_3$}
\psfrag{(a)}{(a)}
\psfrag{(b)}{(b)}
\psfrag{+}{$+$}
\psfrag{-}{$-$}

\includegraphics[width =0.8\textwidth]{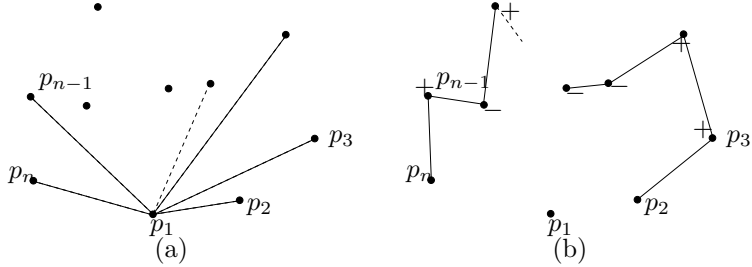}
\caption{Labeling elements of  $P_n.$}
\label{my_proof_03}

\end{figure}

We label every $p \in P_n\setminus \{p_1\}$ according to the angle $\theta$ between the line $y = y_1$ and the line $\overline{p_1 p}.$ The point $p$ will be labeled $p_{i+1}$ if it has the $i$--th smallest angle $\theta,$ see Figure \ref{my_proof_03}(a). For $i = 3,4,...,n-1,$ we say $p_i$ is negative, labeled 
$-,$ if $p_i \in Conv( \{p_1,p_{i-1},p_{i+1}\})^o.$  Otherwise we say $p_i$ is positive, labeled $+.$ See Figure \ref{my_proof_03}(b).

Let ${A}$ and ${B}$ be the subsets of  $P_n$ containing all positive and negative elements respectively. We divide ${A}$ into subsets of consecutive points as follows:

If $p_3 \in {A},$ we define $A_1 = \{p_3,...,p_{3+r-1}\},$ as the subset with $r$ consecutive positive points, where $p_{r+3}$ is negative or $p_{r+3} = p_n.$ If $p_3 \not \in {A}$ then $A_1 = \emptyset.$

Suppose that $p_{n-1} \in {A}.$ For $i \geq 2$ let $A_i' = {A} \setminus \left(A_1 \cup  ... \cup A_{i-1} \right),$ and let
$A_i = \{p_j,p_{j+1},...,p_{j+r-1}\},$ where $r\geq 1,$  $p_j$ has the smallest index in $A_i',$ and $p_{j+r}$ is negative or 
$p_{j+r} = p_{n}.$ Let $k$ be the number of such $A_i$ sets obtained.

If $p_{n-1} \not \in {A},$ we make $A_{k-1}$ the block containing the element in ${A}$ with the highest label, and then  $A_k = \emptyset.$

In an analogous way we partition ${B}$ into $B_1$, $B_2$, ..., $B_{k-1}.$ Let $V$ be the polygon with vertex set ${A}\cup \{p_1,p_2,p_n\},$ and let $U'$ be the set of at most $c-2$ regions $Conv(P_n) \setminus V.$ We call $U$  the vertex set of $U'.$ We obtain a minimal convex decomposition $\Gamma$ of $P_n$ induced by polygons in $V$ and $U$ in the following way:

(1) If $A_j=\{p_i,p_{i+1},...,p_{i+r-1}\}$, we make ${\cal A}_j = A_j \cup \{p_1,p_{i-1},p_{i+r}\}$. ${\cal A}_j$ is the vertex set of an empty convex $(|A_j|+3)$--gon. In case that $A_1 = \emptyset$ (or $A_k = \emptyset$) then ${\cal A}_1 = \{p_1,p_{2},p_{3}\}$
(${\cal A}_k = \{p_1,p_{n-1},p_{n}\}$). There are $k$ of such polygons.

(2) If $B_j=\{p_i,p_{i+1},...,p_{i+r-1}\},$ we make ${\cal B}_j = B_j\cup \{p_{i-1},p_{i+r}\}$. ${\cal B}_j$ is the vertex set of an empty convex $(|B_j|+2)$--gon. There are $k-1$ of them.

(3) Every $B_j = \{p_i,p_{i+1},...,p_{i+r-1}\}$ induces $|B_j|-1$ triangles $\tr p_1 p_m p_{m+1},$ for $m = i,i+1,...,i+r-2.$ There are
$|B_1| - 1 + |B_2| - 1 + ... + |B_{k-1}|-1$ of these triangles. Let $T_B$ be the set of them.

(4) $U'$ can be subdivided in $|A_1|+|A_2|+...+|A_k| - (c-3)$ triangles with vertices in $U$ satisfying (C1) and (C2). Make $T_U$ the set of such triangles.

\begin{figure}[h]
\centering

\psfrag{A1}{$A_1$}
\psfrag{A2}{$A_2$}
\psfrag{A3}{$A_3$}
\psfrag{A4}{$A_4$}

\psfrag{B1}{$B_1$}
\psfrag{B2}{$B_2$}
\psfrag{B3}{$B_3$}

\psfrag{U}{$U$}

\psfrag{p1}{$p_1$}
\psfrag{p2}{$p_2$}
\psfrag{pn}{$p_n$}

\includegraphics[width =0.6\textwidth]{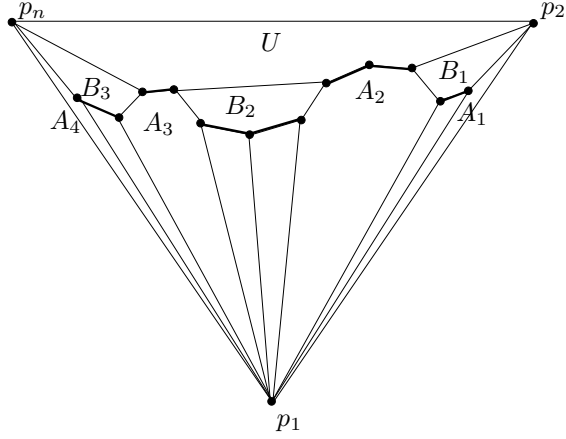}
\caption{$P_{15}$ and $\Gamma$ described above. Here $k = 4$, so $|\Gamma|  = n+k-3 = 16.$}
\label{convdescAB}

\end{figure}

Hence,  $\Gamma = \cup_{i} ({\cal A}_i \cup {\cal B}_i) \cup T_U \cup T_B$ is a convex decomposition of $P_n.$ See Figure \ref{convdescAB}. We have that $|\Gamma| = k + k-1 + |T_B|+|T_U|$ {\em i.e.}

\begin{equation}\label{cardinalidad}
|\Gamma| = n+k-c.
\end{equation}

\subsection{Convex decomposition with at most $\frac{10n}{7}-c$ elements}

We proceed now to show that every collection $P_n$ with $c$ vertices in $Conv(P_n)$ has a convex decomposition $\Gamma$ such that $|\Gamma| \leq \frac{10}{7}n-c.$
We use the following notation: If in a given collection $P_n$ we find that $p_3,$ $p_5,$ $p_7,$ ... are negative and $p_4,$ $p_6,$ $p_8,$ ... are positive, we say that $P_n$ is a $\pm$ set. Next result is for 
$\pm$ sets.

\begin{lemma}\label{lemamine}
Let $P_n$ be a $\pm$ set. Then $P_n$ has a convex decomposition
$\Gamma$ with $\frac{4n}{3}-c$ elements, where $c$ is the number of vertices in $Conv(P_n)$.
\end{lemma}

\begin{figure}[h]
\centering

\psfrag{p1}{$p_1$}
\psfrag{pi+1}{$p_{i+1}$}
\psfrag{pi+2}{$p_{i+2}$}
\psfrag{pi+3}{$p_{i+3}$}
\psfrag{pi+4}{$p_{i+4}$}
\psfrag{pi+5}{$p_{i+5}$}
\psfrag{pi+6}{$p_{i+6}$}
\psfrag{pi}{$p_{i}$}

\psfrag{t1}{$t_1$}
\psfrag{t2}{$t_2$}
\psfrag{t3}{$t_3$}
\psfrag{t4}{$t_4$}
\psfrag{t5}{$t_5$}
\psfrag{t6}{$t_6$}
\psfrag{t7}{$t_7$}
\psfrag{t8}{$t_8$}
\psfrag{t9}{$t_9$}

\includegraphics[width =0.6\textwidth]{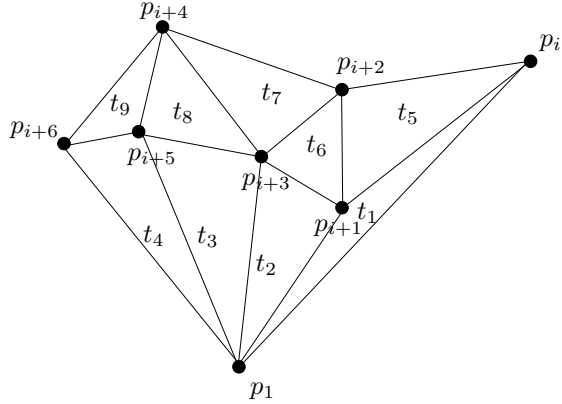}
\caption{$Q_i$ and its convex decomposition $T_i.$}
\label{qidelta}

\end{figure}

\dem For $i = 2,8,...,n-6,$ we make $Q_i = \{p_1, p_{i}, p_{i+1}, ..., p_{i+6}\},$ and let $T_i=\{t_1,t_2,...,t_9\}$ be a set of triangles with vertices in $Q_i$  such that
$t_1 = \tr  p_1 p_{i} p_{i+1}$,
$t_2 = \tr  p_1 p_{i+1} p_{i+3}$,
$t_3 = \tr  p_1 p_{i+3} p_{i+5}$,
$t_4 = \tr  p_1 p_{i+5} p_{i+6}$,
$t_5 = \tr  p_i p_{i+1} p_{i+2}$,
$t_6 = \tr  p_{i+1} p_{i+2} p_{i+3}$,
$t_7 = \tr  p_{i+2} p_{i+3} p_{i+4}$,
$t_8 = \tr  p_{i+3} p_{i+4} p_{i+5}$ and
$t_9 = \tr  p_{i+4} p_{i+5} p_{i+6},$
as shown in Figure~\ref{qidelta}.

We obtain a set $\Gamma_i$ of convex polygons,  joining elements in $T_i,$ to get a minimal convex decomposition of $P_n.$ We make a final modification on positive and negative points as follows: Given 3 consecutive points  labeled $+,$ $p_i,$ $p_j$ and $p_k$ ($i<j<k$),  if
$p_j \in Conv(\{p_1,p_i, p_k\})^o$ we label $p_j$ as $+ -$, otherwise we label $p_j$ as $+ +.$  Analogously we modify labels $-$ to $- -$ and $- +.$

We proceed now to make case analysis over the labels in $p_{i+2},$ $p_{i+3}$ and $p_{i+4}.$ Let
$\ell$ be the line containing $p_{i+1}$ and $p_{i+3},$ make ${\cal D}$ the open half
plane bounded by $\ell$ containing $p_1,$ and ${\cal U} = \mathbb{R}^2 \setminus ({\cal D} \cup \{\ell\}).$ Given two polygons
$\alpha$ and $\beta$ sharing an edge $e,$ we denote as $\alpha \uplus \beta$ the polygon $\alpha \cup (\beta -\{e\}).$

\begin{figure}[h]
\centering

\psfrag{p1}{$p_1$}
\psfrag{pi}{$p_i$}
\psfrag{pi+1}{$p_{i+1}$}
\psfrag{pi+2}{$p_{i+2}$}
\psfrag{pi+3}{$p_{i+3}$}
\psfrag{pi+4}{$p_{i+4}$}
\psfrag{pi+5}{$p_{i+5}$}
\psfrag{pi+6}{$p_{i+6}$}

\psfrag{(a)}{(a)}
\psfrag{(b)}{(b)}

\includegraphics[width = 0.9\textwidth]{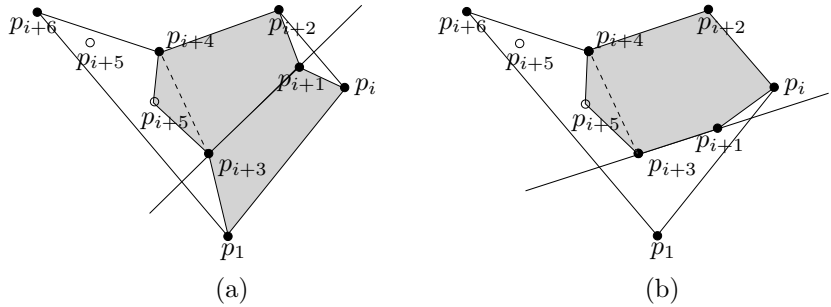}
\caption{$p_{i+2},$ $p_{i+4}$ and $p_{i+3}$ being $+ +,$ $+ -$ and $- -$ respectively.}
\label{casoA1}

\end{figure}

\noindent {\bf Case (a).} $p_{i+2}$ and $p_{i+4}$ have label $+ +$ and $+ -$ respectively. We have:

{\em Subcase 1.} Suppose that $p_{i+3}$ is $- -.$  If $p_{i}\in {\cal D}$ and the pentagon $P = t_6 \uplus t_7 \uplus t_8$ is convex, then
$\Gamma_i =\{t_1 \uplus t_2,t_3,t_4,t_5,P, t_9\}.$ If $P$ is not convex,
$\Gamma_i =\{t_1 \uplus t_2,t_3 \uplus t_8, t_4,t_5,t_6 \uplus t_7, t_9\}.$
See Figure \ref{casoA1}(a).

If $p_i \in {\cal U},$ and the hexagon $H = t_5 \uplus t_6 \uplus t_7 \uplus t_8$ is convex,
$\Gamma_i =\{t_1, t_2, t_3, t_4, H, t_9\}.$ If $H$ is not convex,
$\Gamma_i =\{t_1,t_2,t_3 \uplus t_8,t_4,t_5 \uplus t_6 \uplus t_7, t_9\}.$
See Figure \ref{casoA1}(b).

\begin{figure}[h]
\centering

\psfrag{p1}{$p_1$}
\psfrag{pi}{$p_i$}
\psfrag{pi+1}{$p_{i+1}$}
\psfrag{pi+2}{$p_{i+2}$}
\psfrag{pi+3}{$p_{i+3}$}
\psfrag{pi+4}{$p_{i+4}$}
\psfrag{pi+5}{$p_{i+5}$}
\psfrag{pi+6}{$p_{i+6}$}

\psfrag{(a)}{(a)}
\psfrag{(b)}{(b)}
\psfrag{(c)}{(c)}
\psfrag{(d)}{(d)}

\includegraphics[width = 0.9\textwidth]{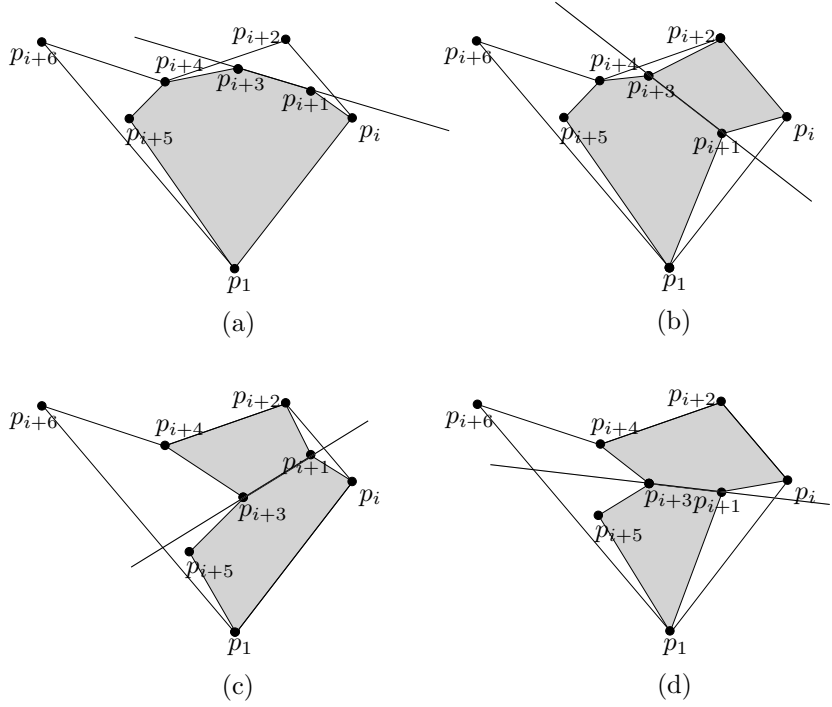}
\caption{Polygons we find when $p_{i+3}$ is $-+$ and $p_{i+2}$ and $p_{i+4}$ are $++$ and $+-$ respectively. We find same polygons if both of $p_{i+2}$ and $p_{i+4}$ are labeled $++.$}
\label{casoA2}

\end{figure}

{\em Subcase 2.} Suppose that $p_{i+3}$ is $- +.$ If $p_{i}$ and $p_{i+4}$ are in ${\cal D},$ then make
$H=t_1 \uplus t_2 \uplus t_3 \uplus t_8$ and $\Gamma_i =\{H,t_4,t_5,t_6,t_7, t_9\}$ (see Figure \ref{casoA2}(a)).
Now, if $p_i \in {\cal U}$ (and $p_{i+4} \in {\cal D}$) $H$ is missing $t_1,$ so $\Gamma_i =\{t_1,t_2 \uplus t_3 \uplus t_8,t_5 \uplus t_6,t_4,t_7, t_9\}$
(see Figure \ref{casoA2}(b)).
On the other hand if $p_{i+4} \in {\cal U}$
(and $p_{i} \in {\cal D}$), $H$ is missing $t_8,$ so $\Gamma_i =\{t_1 \uplus t_2 \uplus t_3,t_4,t_5,t_6 \uplus t_7,t_8, t_9\}$
(see Figure \ref{casoA2}(c)).
Finally if $p_i, p_{i+2}, p_{i+4} \in {\cal U},$ $\Gamma_i =\{t_1,t_2 \uplus t_3,t_4,t_5 \uplus t_6 \uplus t_7,t_8, t_9\}$
(see Figure \ref{casoA2}(d)).

\noindent {\bf Case (b).} Both $p_{i+2}$ and $p_{i+4}$ have label $+ -.$ Observe that $\{p_{i},p_{i+2},p_{i+4},p_{i+6}\}$ is the set of vertices of a convex quadrilateral $q,$ so we make $\Gamma_i = \{   t_1, t_2 \uplus t_6, t_3 \uplus t_8, t_4, t_5, t_7, t_9, q \},$ $U = U\setminus \{p_{i+2},p_{i+4}\}$ and $U' = U'\setminus q.$

\begin{figure}[h]
\centering

\psfrag{p1}{$p_1$}
\psfrag{pi}{$p_i$}
\psfrag{pi+1}{$p_{i+1}$}
\psfrag{pi+2}{$p_{i+2}$}
\psfrag{pi+3}{$p_{i+3}$}
\psfrag{pi+4}{$p_{i+4}$}
\psfrag{pi+5}{$p_{i+5}$}
\psfrag{pi+6}{$p_{i+6}$}

\psfrag{(a)}{(a)}
\psfrag{(b)}{(b)}

\includegraphics[width =0.9\textwidth]{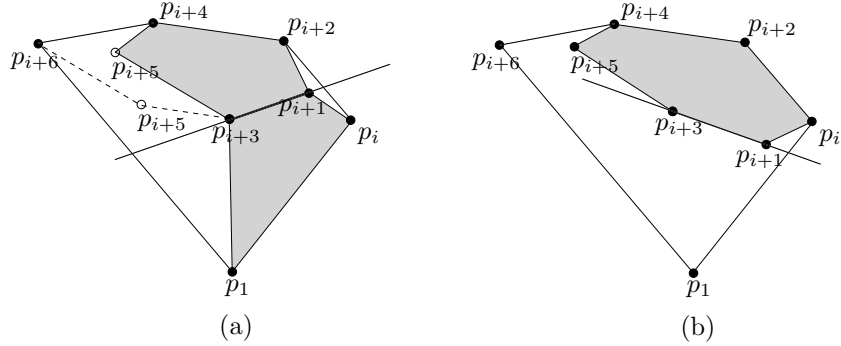}
\caption{Polygons we find when $p_{i+3}$ is $- -,$ and $p_{i+2}$ and $p_{i+4}$ are both $++.$}
\label{casoC1}

\end{figure}

\noindent {\bf Case (c).} $p_{i+2}$ and $p_{i+4}$ both have label $+ +.$ We are making a similar analysis as in Case (a): Suppose that $p_{i+3}$ is $- -.$ If $p_i \in {\cal D}$ and hexagon $H = t_5 \uplus t_6 \uplus t_7 \uplus t_8$ is convex, $\Gamma_i =\{t_1,t_2,t_3,t_4,H, t_9\}.$ If $H$ is not convex, we make
$\Gamma_i =\{t_1 \uplus t_2,t_3,t_4,t_5,t_6 \uplus t_7 \uplus t_8, t_9\}.$  See Figure \ref{casoC1}(a).

If $p_{i}\in {\cal U},$ $H$ is always convex, so $\Gamma_i =\{t_1,t_2,t_3,t_4,H, t_9\}.$ See Figure \ref{casoC1}(b).

When $p_{i+3}$ is $- +,$ $\Gamma_i$ has the same polygons as in subcase 2 of Case (a). And if $p_{i+2}$ and $p_{i+4}$ have label $+ -$ and $+ +$ respectively, we obtain $\Gamma_i$ analogously as in Case (a).

%%%%%%%%%%%%%%%%%%%%%%%%%%%%%%%%%%%%%%

Lets make $R_i = \gamma_i \uplus \gamma_{i+1}$ where $\gamma_i$ is the polygon containing $t_4$ in $Q_i,$ and $\gamma_{i+1}$ is the polygon containing $t_1$ in $Q_{i+1},$ and let $b$ be the number of $Q_i$ sets as in case (b). We obtain a minimal convex decomposition of $P_n$ by finding 
$\Gamma_2,$ $\Gamma_8,$ ... ,$\Gamma_{n-6},$ obtaining the $\frac{n}{2}-c-2b$ triangles in $T_U,$ and getting $R_i$ by removing edges $p_1p_{i},$  for $i = 8, 14, 20, ... ,n-6.$

So $\Gamma$ is such that $|\Gamma| = \left(6\frac{n}{6} + 2b \right) + \left( \frac{n}{2}-c-2b  \right) - \left(\frac{n}{6} \right) = \frac{4n}{3} - c.$ \fin

We have the following observation.

\noindent {\bf Observation 1.} 
Let $\gamma$ be the vertex set of a convex polygon, and let $p$ be a point in $Conv(\gamma)^o,$ then $\gamma \cup p$ has a minimal convex decomposition with 3 elements.

We proceed now to prove our main theorem.

\begin{theorem}
Let $P_n$ be an $n$--point set on the plane in general position. Then $P_n$ has a minimal convex decomposition with at most $\frac{10}{7}n-c$ elements.
\end{theorem}

\dem Let $k$ be the number of polygons ${\cal A}_i$ described above. If $k \leq \frac{3n}{7},$ we apply Equation~(\ref{cardinalidad}) to find a convex decomposition with
$n + k - c \leq n + \frac{3n}{7}-c$ elements. If $k = \frac{n}{2}$, $P_n$ is a $\pm$ set, and it has a convex decomposition with $\frac{4n}{3}-c$ elements, 
by Lemma~\ref{lemamine}.

In case that  $\frac{3n}{7}< k < \frac{n}{2},$ we consider every ${\cal A}_i.$ Let $I = {\cal A}_i \cap Conv(P_n)^o.$ 
If $I={\cal A}_i$ let $q_i$ be the element in ${\cal A}_i$ with the highest coordinate $y$ (if there are 2 points with this coordinate, we make $q_i$ the element having the greatest $x$ coordinate of them).

\begin{figure}[h]
\centering

\psfrag{A1}{$A_1$}
\psfrag{A2}{$A_2$}
\psfrag{A3}{$A_3$}
\psfrag{A4}{$A_4$}

\psfrag{B1}{$B_1$}
\psfrag{B2}{$B_2$}
\psfrag{B3}{$B_3$}

\psfrag{p1}{$p_1$}
\psfrag{p2}{$p_2$}
\psfrag{pn}{$p_n$}

\includegraphics[width =0.9\textwidth]{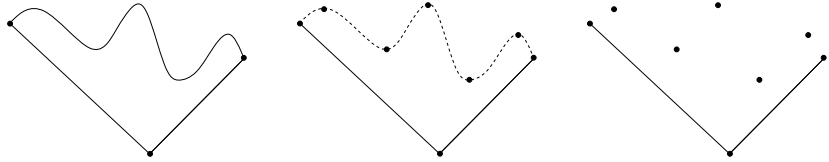}
\caption{$P_n$ and its collection $\pm$ associated.}
\label{coleccionPM}
\end{figure}

If $I \not = {\cal A}_i,$ let $q_i$ be the element with the highest label in ${\cal A}_i - I,$ and make $r_i$ in $B_i$ the element with minimum $y$ coordinate, if there are 2 with the same coordinate, we make $r_i$ the element with maximum $x$ coordinate. See Figure~\ref{coleccionPM}.

We make $P' = \{q_1, r_1, q_2, r_2, ... , q_{k-1}, r_{k-1}, q_k\} \cup \{p_1,p_2,p_n\}.$ $P'$ is a $\pm$ set with $2k+2$ elements. By Lemma~\ref{lemamine}, $P'$ has a convex decomposition $\Gamma'$ with  $\frac{4}{3} (2k+2) -c$ elements. Let $S$ be the set  $P_n - P',$ where $|S|= n - 2k - 2.$ By Observation 1, we find that every element in $S$ when is added increases in 2 the number of polygons, so $P'$ and $S$ induce a minimal convex decomposition $\Gamma$ of $P'\cup S=P_n$ with $\frac{4}{3} (2k+2) -c + 2|S|$ elements. Substituting $|S|$ we have that $|\Gamma| = 2n-\frac{4}{3}k-c-\frac{4}{3}.$ Using the fact that
$k \geq \frac{3n}{7}$ we obtain that $\Gamma$ is such that $|\Gamma| \leq \frac{10n}{7}-c.$ \fin

\section{Concluding remarks}

Analogously to triangulation of $P_n,$ we can define {\em convex quadrangulation}. It would be interesting characterizing $n$--point sets that accept a convex qua\-dran\-gu\-la\-tion.

\end{document}